\documentclass[3p,times,procedia]{elsarticle}
\usepackage{nupha_ecrc}

\usepackage{mathtools}

\newcommand{\eq}[1]{\begin{align} #1 \end{align}}

\volume{00}

\firstpage{1}

\journalname{Nuclear Physics A}

\runauth{}

\jid{nupha}

\jnltitlelogo{Nuclear Physics A}

\usepackage{amssymb}

\usepackage[figuresright]{rotating}

\begin{document}

\begin{frontmatter}


\title{Net-baryon fluctuations measured with ALICE at the CERN LHC}

\author[1,2,3,4]{Anar Rustamov (for the ALICE Collaboration)}

\address[1]{Azerbaijan National Academy of Sciences, Institute of Physics, Baku, Azerbaijan}
\address[2]{Physikalisches Institut, Universit\"{a}t Heidelberg, Heidelberg, Germany}
\address[3]{Extreme Matter Institute EMMI, GSI, Darmstadt, Germany}
\address[4]{National Nuclear Research Center, Baku, Azerbaijan}

\begin{abstract}

First experimental results are presented on event-by-event net-proton fluctuation measurements in Pb- Pb collisions at $\sqrt{s_{NN}}$ =
2.76 TeV, recorded by the ALICE detector at the CERN LHC. The ALICE detector is well suited for such studies due to its excellent particle identification capabilities and large acceptance, which is 
crucial for fluctuation analysis.
The studies are focussed on second order cumulants, but the analysis technique used is more general
and will be applied, in the near future, also to higher order cumulants. 
\end{abstract}

\begin{keyword}
Quark-gluon plasma \sep Fluctuations \sep Conservation laws

\end{keyword}

\end{frontmatter}

\section{Introduction}
\label{lintroduction}
Phase transitions in strongly interacting systems can be addressed by investigating the
response of the system to external perturbations via fluctuations of conserved charges. In particular, one can look for critical fluctuations at vanishing baryon chemical potential as reported in~\cite{LQCD1, Redlich1}. 
However, fluctuations of conserved charges are predicted in the Grand Canonical Ensemble (GCE)
formulation of thermodynamics~\cite{StatLandau}. In this formulation, the net-baryon
number is not conserved in each micro-state, hence it fluctuates. In order to
compare theoretical calculations within GCE, such as the Hadron Resonance Gas (HRG) model~\cite{HRG} and
Lattice QCD (LQCD)~\cite{LQCD1}, to experimental results, the requirements of GCE have to be
achieved in experiments. This is typically done by analysing the
experimental data in a finite acceptance by imposing cuts on
rapidity and/or  transverse momentum of detected
particles. However, if the selected acceptance window is too small,
the possible dynamical correlations we are after will also be strongly
reduced~\cite{KochFluct} and consequently, net-baryons will be
distributed according to the difference of two independent Poisson
distributions~\cite{ourModel}.  We remind that for the Poisson distribution, all its cumulants are equal to
its mean. The probability distribution of the difference $X_{1}$ -
$X_{2}$ of two random variables, each generated from statistically
independent Poisson distributions, is called the Skellam distribution.
According to the additivity of cumulants, the cumulants of the Skellam
distribution will then be $\kappa_{n}(Skellam)  = \left<X_{1}\right> + (-1)^{n}\left<X_{2}\right>$, where $\left<X_{1}\right>$ and $\left<X_{2}\right>$ are mean values of  $X_{1}$ and $X_{2}$ respectively.

On the other hand,  the increase of the acceptance will enlarge significance of correlations  due
to baryon number conservation. In order to be more sensitive to dynamical fluctuations, the better approach is to study the fluctuation of conserved
charges in a larger acceptance and subtract the correlation part caused by the global conservation laws.
This is actually the appropriate way to address the fluctuation physics when both, trivial and dynamical
fluctuations are close to Poisson probability distributions. To proceed further we provide the necessary definitions. 
The first and second cumulants of net-baryon $\Delta n_{B} = n_{B} - n_{\bar{B}}$ distribution is defined as:
\eq{
\label{kumulants_definition}
  &\kappa_{1}(\Delta n_{B}) = \left<\Delta n_{B}\right> , \\
  &\kappa_{2} (\Delta n_{B}) = \left<\Delta n_{B}^{2}\right> - \left< \Delta n_{B} \right> ^{2}. }
The second cumulant can be represented  as a sum of corresponding cumulants for
single baryons plus the correlation term for joint probability distributions of baryons and antibaryons
\begin{equation}
\kappa_{2}(\Delta n_{B})= \kappa_{2}(n_{B})+\kappa_{2}(n_{\bar{B}}) - 2\left(\left<n_{B}n_{\bar{B}}\right> -
\left<n_{B}\right>\left<n_{\bar{B}}\right>\right),
\label{sec_cum_2}
\end{equation}
Eq.~\ref{sec_cum_2} shows
that, in the case of missing correlations between baryons and
antibaryons, the second cumulant of  net-baryons  is exactly
equal to the sum of the corresponding second cumulants for baryons
and antibaryons.

A correlation can emerge, in addition to that originating from 
critical fluctuations, from conservation laws. If the experimental acceptance is large enough the
correlation term in Eq.~\ref{sec_cum_2} acquires a finite value.
Hence we expect that the second cumulant of net-baryons becomes
smaller than the sum of baryons and antibaryons.  

We note that in this work  net-protons are used as a proxy for net-baryons, which is justified at LHC energies~\cite{Kitazawa}.

\section{The experimental data and analysis method}
The results presented are obtained by analysing about 13
million minimum-bias Pb-Pb events at $\sqrt{s_{NN}}$ = 2.76 TeV recorded by the ALICE detector~\cite{ALICE} in the year 2010. The detectors used are the Time Projection Chamber (TPC) for tracking and particle identification and the Inner
Tracking System (ITS) for precise vertex determination. Both devices are located in the central barrel of the experiment and operate inside a large solenoidal magnet with $B=0.5$ T. 
Two forward scintillator hodoscopes V$0$ are located on either side of the interaction point and cover pseudorapidity intervals $2.8 < \eta < 5.1$ and $-3.7 < \eta < -1.7$. The minimum-bias trigger condition is defined by the coincidence of hits in both V$0$ detectors.  The event centrality is selected based  on the signal amplitudes in the V$0$ detectors. 
The phase space coverage is restricted to $0.6 < p[$GeV/c$] < 1.5$, and $|\eta| < 0.8$.  Moreover, differential analysis is provided as function of $\Delta\eta$. High precision particle identification in a wide momentum range is achieved by correlating the measured particle momentum with its specific energy loss $dE/dx$ in the gas volume of  TPC.  The cumulants of net-protons are reconstructed using a novel approach, the Identity Method, which overcomes incomplete particle identification event-by-event, caused by the overlapping $dE/dx$ distributions~\cite{Iden1, Iden2, Iden3, Iden4, Mesut}. This allows for event-by-event fluctuation analysis with high PID and tracking efficiencies, which amounts to about 80$\%$ for protons, almost independent of the collisions centrality.  The latter is important because small efficiencies also reduce dynamical fluctuations of interest.  As an input the method uses only the probabilities for each track in a given event of being a proton, kaon, pion or electron, which are calculated by exploiting fitting $dE/dx$ distributions within the fine bins of track momentum and pseudorapidity~\cite{Iden4, Mesut}.

\section{Results and discussions}
\label{lresults}
\begin{figure}[htb]
\centering
 \includegraphics[width=0.4\linewidth,clip=true]{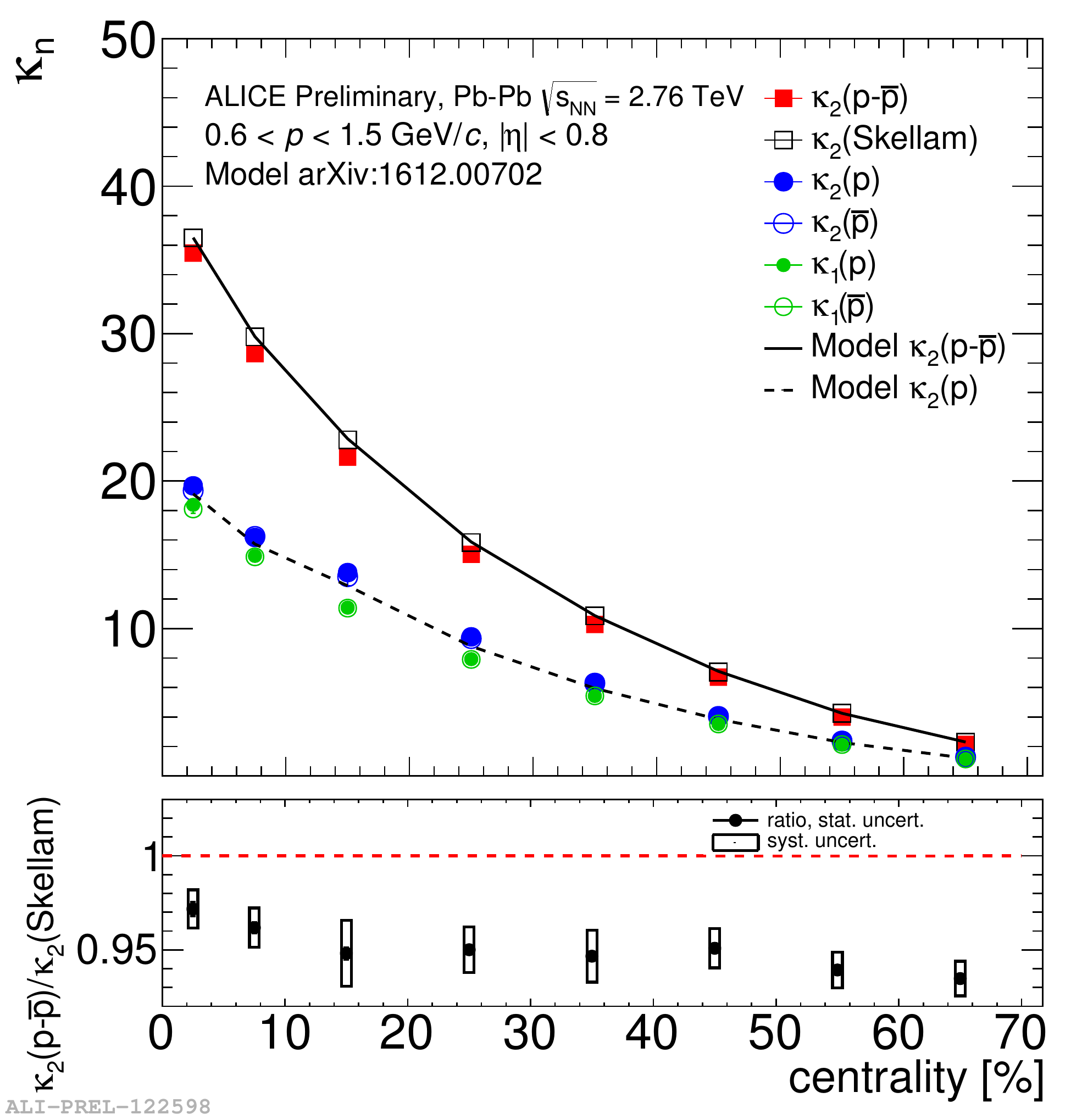}
 \includegraphics[width=0.5\linewidth,clip=true]{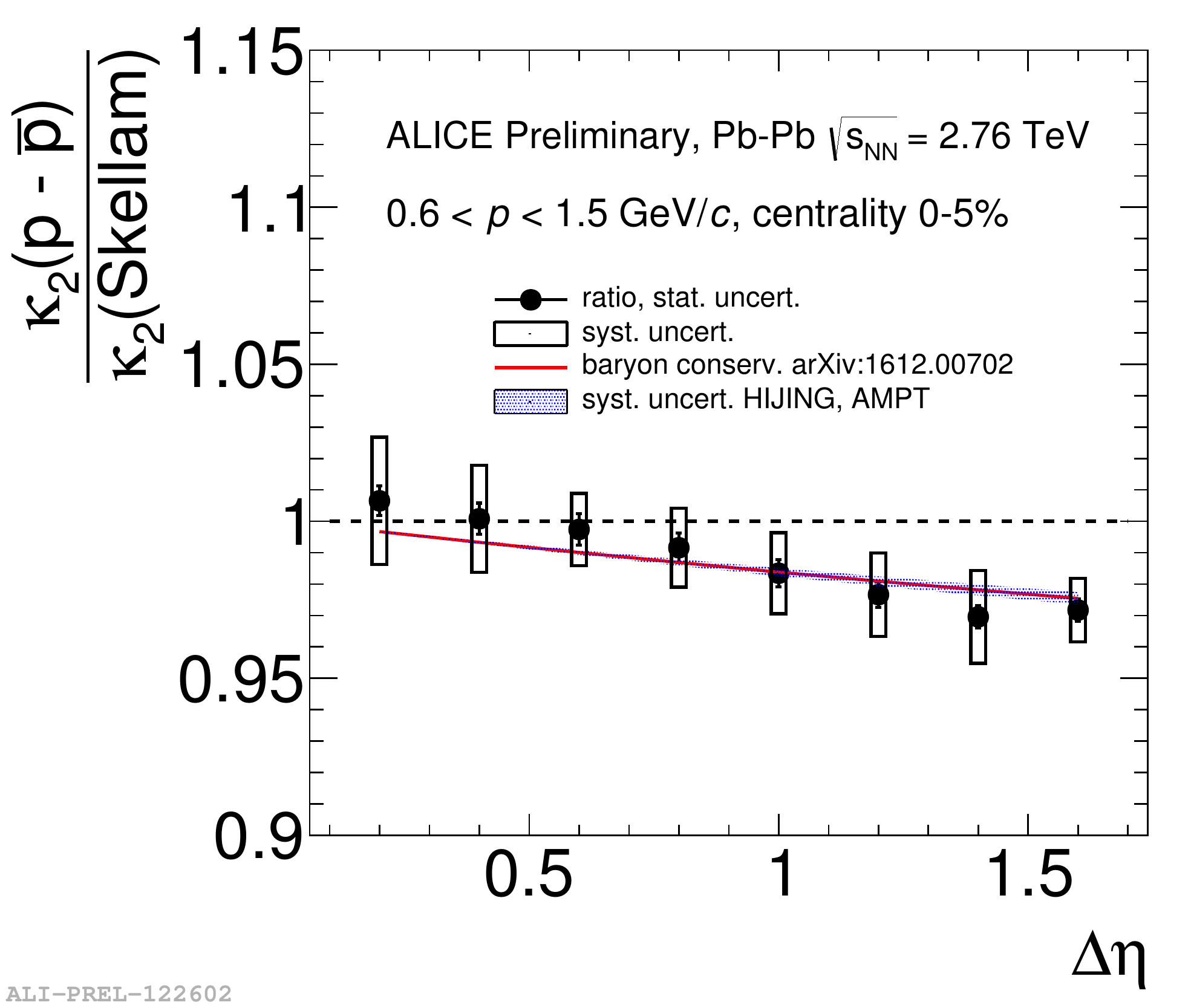} 
 \caption{Left panel: Experimentally measured second cumulants of net-proton distributions (red solid boxes) compared to Skellam baseline (open squares).
 The second cumulants of single proton and antiproton distributions are presented with the filled and open blue circles, correspondingly. 
 The green solid and open circles represent first cumulants of protons and antiprotons respectively, which are hardly distinguishable because of the nearly equal mean numbers of protons and antiprotons. The model predictions with the underlying independent Poisson distributions for protons and antiprotons
 are depicted with the solid (net-protons) and dashed (protons) lines. In the bottom panel the ratio of the experimentally measured second cumulants of net-protons to the Skellam baseline is presented. Right panel: Pseudorapidity   dependence of the normalised second cumulants of net-protons. The red solid line shows the effect of the baryon number conservation. 
 }
\label{fnet-prootns}
\end{figure}

In the left panel of Fig.~\ref{fnet-prootns} the centrality dependence of the  second cumulants of net-protons, represented with the solid red boxes, is compared to the Skellam baseline depicted with the open boxes. 
We note that the Skellam baseline is calculated as the sum of the mean numbers (first cumulants) of protons (solid green circles) and antiprotons (open green circles) in the corresponding centrality class. Below we demonstrate the validity of this approach for the baseline calculation. As seen from the bottom panel of Fig.~\ref{fnet-prootns},  deviations of the second cumulants of net-protons from the Skellam baseline are observed for all centrality classes.  In order to shed light on these observations we have reconstructed second cumulants of single proton and antiproton distributions which are depicted in in the left panel of Fig.~\ref{fnet-prootns} with the solid and open blue circles, respectively. We observe significant differences between second and first cumulants of single protons and anti-protons. The latter however does not necessarily indicate deviation of single proton and antiproton distributions from the underlying Poisson baseline. Indeed, within the recently proposed model it was demonstrated that dynamical fluctuations are significantly modified by the unavoidable fluctuations of participant nucleons~\cite{ourModel} (see also Ref.~\cite{Skokov}).  The model uses several inputs such as mean number of protons and antiprotons and the centrality selection procedure which determines the fluctuations of participants. Using the experimentally measured mean values of protons and antiprotons presented  in Fig.~\ref{fnet-prootns}  and the same centrality selection as used in this analysis  we calculated  second cumulants of protons and net-protons in the presence of participant fluctuations. These results are presented with the dashed and solid lines in the left panel of Fig.~\ref{fnet-prootns} for protons and net-protons respectively. Both calculations  are consistent with the experimentally measured second cumulants of protons and the Skellam distribution, correspondingly. We note that, in the model, particles are produced from the independent Poisson distributions, i.e, the difference between the dashed line and the mean values of protons is completely driven by the participant fluctuations.  We therefore conclude that the observed deviation between the second and first cumulants of protons and antiprotons are stemming from participant fluctuations.  On the other hand, the consistency between the solid line and the Skellam distribution shows that the second cumulants of net-protons at LHC energies are not affected by the participant fluctuations, which justifies the calculation of the Skellam baseline as a sum of the mean numbers of protons and antiprotons.  According to Eq.~\ref{sec_cum_2}  the only reason for the deviation of the experimentally measured second cumulants of net-protons from the corresponding second cumulants of the Skellam distribution can be due to the correlation term.  

For the second cumulants the correlation term arising from the global baryon number conservation depends only on the acceptance factor $\alpha = \left< n_{p}\right> / \left< N_{B}^{4\pi}\right> $ with $\left<n_{p}\right>$ and $\left< N_{B}^{4\pi}\right>$ referring to the mean number of protons
inside the acceptance and the mean number of baryons in the full phase space respectively~\cite{ourModel}:

\begin{equation}
\frac{\kappa_{2}\left(p-\bar{p}\right)}{\kappa_{2}\left(Skellam\right)}=1-\alpha.
\label{conserv2}
\end{equation}
We performed our analysis in 8 different pseudo-rapidity ranges from
$|\eta| < 0.1$ up to $|\eta| < 0.8$ in steps of 0.1. The obtained
results for the second cumulants of net-protons, normalised to the
Skellam baseline, are presented in the right panel of  Fig.~\ref{fnet-prootns}.
For each rapidity range we estimated the acceptance factor
$\alpha$. In doing so, we used
the total number of baryons as measured by the ALICE experiment in the
pseudorapidity range of $|\eta| < 0.5$~\cite{fluct_exp}.  Next, using
HIJING and AMPT simulations, we obtained total number of baryons in
the full phase space. The number of protons, used in the definition of $\alpha$ (cf. Eq.~\ref{conserv2}),
is taken from the current analysis for each
rapidity range.  Finally, using these values of $\alpha$ the red
band in Fig.~\ref{fnet-prootns} is calculated with 
Eq.~\ref{conserv2}. The finite width of the band reflects 
the difference between the two event generators.

As seen from the right panel of  Fig.~\ref{fnet-prootns}, for
pseudorapidity ranges of $|\eta| <0.4$, which corresponds to
$\Delta\eta<0.8$, the experimentally measured net-proton distributions follow a
Skellam distribution. This  agreement is due to the small acceptance as discussed above.
Beyond $\Delta |\eta| > 0.8$ we do observe
deviations from the Skellam distribution. The amount of deviation is
in good agreement with the red band as calculated with 
Eq.~\ref{conserv2}. We hence conclude that this deviation is caused by
global baryon number conservation and that other possible dynamical
fluctuations are not visible in the second cumulants of net-protons.

\section{Conclusions}
In summary, we presented first measurements of net-proton fluctuations from the ALICE experiment at LHC. 
The measured second cumulants of net-protons, which are used as a proxy for net-baryons, are, after accounting for baryon number conservation, in 
agreement with the corresponding second cumulants of the Skellam distribution. We note that LQCD predicts a Skellam behaviour for the second cumulants of net-baryon distributions at a pseudo-critical temperature of about 155 MeV,
which is very close to the freeze-out temperature from the HRG model applied to the ALICE data~\cite{HRG, Karsch}.  Critical behaviour is predicted by LQCD for higher cumulants of net-baryon distributions, which will be the topic of our further investigations~\cite{Karsch}.

\section*{Acknowledgments}
This work is part of and supported by the DFG Collaborative Research
Centre "SFB 1225 (ISOQUANT)".

\end{document}